\documentstyle[12pt,a4]{article}

\newcommand{\my}{\setcounter{equation}{0}}
\begin{document}

\begin{flushright}
\begin{minipage}[t]{4cm}
DPNU-96-08 \\
hep-ph/9602338 \\
February 1996
\end{minipage}
\end{flushright}
\vspace{1cm}
\begin{center}
\Large{{\bf Isospin Breaking Effects in the Anomalous \\
Processes with Vector Mesons}}
\\
\vspace{1cm}
\large{Michio Hashimoto
\footnote{e-mail address: michioh@eken.phys.nagoya-u.ac.jp}
 }
\\
\footnotesize{\it{
Department of Physics Nagoya University, Nagoya 464-01 Japan}}
\end{center}
\vspace{0.8cm}
\begin{abstract}
We introduce isospin breaking terms as well as \(SU(3)\) breaking 
terms to the anomalous \(VVP\) coupling in the hidden local symmetry 
scheme without affecting the low energy theorem on the processes 
such as \(\pi\to 2 \gamma\) 
and \(\gamma\to 3\pi \). It is shown that the predictions from 
these terms coincide successfully with all the experimental data of 
anomalous decays. It is also predicted that the decay widths of 
\(\rho^0 \to \pi^0 \gamma\) and \(\phi \to \eta' \gamma \) are 
\(114\pm 7 {\rm keV \/}\) and \( 0.552 \pm 0.055 {\rm keV\/} \), 
respectively. 
\\

\(PACS\): 12.39.Fe, 12.40.Vv, 13.25.-k, 13.65.+i, 14.40.Aq, 14.65.Bt
\end{abstract}

\section{Introduction}
\my
Anomalous processes involving vector mesons are interesting probes 
to test the effective theories of QCD through the low-energy and 
high-luminosity \(e^+ e^-\) collider experiments in near future.
In particular, the DA\(\Phi\)NE \(\phi\)-factory 
is expected to yield \(2\times 10^{10} \; \phi\)-meson decays per 
year [1], which will provide us with high quality data for decays of 
pseudoscalar and vector mesons in the light quark sector. 
We have improved upper bounds of the branching ratios of 
rare \(\phi\)-decays such as \(\phi\to\rho\gamma,\; 
\phi\to\omega\gamma\), etc. 
and may be able to obtain the branching ratio of 
\(\phi\to\eta'\gamma\) [1] for which only the upper bound is 
known today [2]. Moreover, 
uncertainty of the data on \(\rho^0\to \pi^0 \gamma\) 
will be much reduced [1].

These radiative decays are associated with the flavor anomaly of QCD 
and are described by the Wess-Zumino-Witten(WZW) term[3] 
in the low energy limit. Based on the hidden local symmetry(HLS) 
[4][5] for the vector mesons, 
Fujiwara et al.[6] proposed 
a systematic way to incorporate vector mesons into such 
a chiral Lagrangian with WZW term without affecting 
the low-energy theorem on \(\pi^0 \to 2\gamma \; , \gamma \to 3\pi\) 
etc. Bramon et al.[8] studied extensively the radiative vector meson 
decays by introducing \(SU(3)\) breaking into the anomalous 
Lagrangian of Fujiwara et al.[6]. However, method of Bramon et al. 
is not consistent with the low-energy theorem, 
especially on \(\eta (\eta') \to 2\gamma\), which are 
essentially determined by the WZW term. Thus, if isospin breaking 
effects were introduced through their method, successful low-energy 
theorem on \(\Gamma(\pi^0 \to 2\gamma)\) and 
\(\Gamma(\gamma \to 3\pi)\) would be violated. 
Furthermore, the breaking effects 
are important to account for the difference between 
\(\Gamma(\rho^0 \to \pi^0 \gamma)\) and 
\(\Gamma(\rho^{\pm} \to \pi^{\pm}\gamma) \). 

In this paper, we construct all possible isospin/\(SU(3)\)-broken 
anomalous HLS Lagrangians with the smallest number of derivatives 
in a manner consistent with the low energy theorem 
in contrast to Bramon et al.[8]. 
This is systematically done through spurion method for 
the breaking term. 
It is further assumed that we can neglect direct 
\(VP\gamma\) and \(VP^3\) couplings 
(\(V\) = vector meson, \(P\) = pseudoscalar meson) which are absent 
in the original Lagrangian[6]. Then we find a parameter region 
which is consistent with all the existing data on radiative decays 
of vector mesons. Such a parameter region yields some predictions on 
the decays like \(\Gamma(\rho^0\to\pi^0\gamma), \; 
\Gamma(\phi\to\eta'\gamma), \; \Gamma(\rho\to \pi\pi\pi) \) and 
\(\Gamma(K^*\to K\pi\pi)\).  

The paper is organized as follows : In section 2, a review of 
HLS Lagrangian is given for both non-anomalous and anomalous terms. 
\(SU(3)\) breaking terms are introduced into the non-anomalous 
HLS Lagrangian \`a la Bando et al.[5]. 
In section 3, we construct the most general isospin/\(SU(3)\)-broken 
anomalous Lagrangians with the lowest derivatives 
in a way consistent with the low energy theorem. 
In section 4, the phenomenological analysis of these Lagrangians 
will be successfully done for the radiative decays of vector mesons. 
In section 5, we make the analysis for the hadronic anomalous decays. 
Section 6 is devoted to summary. 

\section{Hidden Local Symmetry}
\my
Here we give a brief review of HLS approach[7]. 
A key observation is that the non-linear sigma model based 
on the manifold \(U(3)_L \times U(3)_R/U(3)_V\) is gauge equivalent 
to another model having a symmetry 
\([U(3)_L \times U(3)_R ]_{{\rm global\/}} \times 
[U(3)_V]_{{\rm local\/}} \). Vector mesons are introduced as 
the gauge fields of a hidden local symmetry 
\([U(3)_V]_{{\rm local\/}}\). The photon field is introduced through 
gauging \([U(3)_L \times U(3)_R]_{{\rm global\/}} \). 

The HLS Lagrangian is given by [4], [5]: 
\begin{eqnarray}
{\cal L} &=& {\cal L}_A + a {\cal L}_V + {\cal L}_{{\rm gauge\/}}, 
\label {HLS} \\
{\cal L}_A &=& -\frac{f_{\pi}^2}{8}{\rm tr\/} (D_{\mu}\xi_L \cdot 
\xi_L^{\dagger} - D_{\mu}\xi_R \cdot \xi_R^{\dagger})^2 , 
\label{2-2} \\
{\cal L}_V &=& -\frac{f_{\pi}^2}{8} {\rm tr\/} (D_{\mu}\xi_L \cdot 
\xi_L^{\dagger}+ D_{\mu}\xi_R \cdot \xi_R^{\dagger})^2 , \label{2-3} 
\end{eqnarray}
where \(f_{\pi} \simeq 131 {\rm MeV\/}\) is the decay constant of 
pseudoscalar mesons ,\(D_{\mu} \xi_{L,R} \equiv ( 
\partial_{\mu} - igV_{\mu} )\xi_{L,R} + ie\xi_{L,R} 
Q \cdot B_{\mu}\) , with 
\( Q = {\rm diag\/} \left( \frac{2}{3},-\frac{1}{3},-\frac{1}{3} 
\right) \), and \(V_{\mu}\) and \(B_{\mu}\) 
being the vector mesons and the photon 
fields, respectively, and \({\cal L}_{{\rm gauge\/}}\) 
is the kinetic terms of \(V_{\mu}\) and \(B_{\mu}\). Here 
\(g\), \(e\), and \(a\) are respectively the hidden gauge coupling, 
the electron charge and a free parameter not determined by 
the symmetry considerations alone.

The fields \(\xi_{L,R}\) and \(V_{\mu}\) transform as follows; 
\begin{eqnarray}
\xi_{L,R}(x) &\to& \xi_{L,R}'(x)=h(x)\xi_{L,R}(x)
g^{\dagger}_{L,R}(x) \; , \\
V_{\mu}(x) &\to& V_{\mu}'(x)=h(x)V_{\mu}(x)h^{\dagger}(x)+ih(x)
\partial_{\mu}h^{\dagger}(x) \; , 
\end{eqnarray}
where \(h(x)\in[U(3)_V]_{{\rm local\/}}, \; 
g_{L,R}(x)\in[U(3)_{L,R}]_{{\rm global\/}}\). 
To do a phenomenological analysis, we take unitary gauge:
\begin{eqnarray}
\xi_R &=& \xi_L^{\dagger}=e^{\frac{iP}{f_\pi}} ,  \\
P &=& \left( 
\begin{array}{ccc}
\frac{\pi^0}{\sqrt{2}}+\frac{\eta}{\sqrt{3}}+\frac{\eta'}{\sqrt{6}} 
& \pi^+ & K^+ \\
\pi^- & -\frac{\pi^0}{\sqrt{2}}+\frac{\eta}{\sqrt{3}}+
\frac{\eta'}{\sqrt{6}} & K^0 \\
K^- & \bar{K}^0 & -\frac{\eta}{\sqrt{3}}+\sqrt{\frac{2}{3}}\eta'
\end{array}
\right) ,  \\
V &=& \left( 
\begin{array}{ccc}
\frac{\rho^0}{\sqrt{2}}+\frac{\omega}{\sqrt{2}} & \rho^+ & K^{*+} \\
\rho^- & -\frac{\rho^0}{\sqrt{2}}+\frac{\omega}{\sqrt{2}} & K^{*0} \\
K^{*-} & \bar{K}^{*0} & \phi
\end{array}
\right) , 
\end{eqnarray}
where we assumed that \(\eta_1\)-\(\eta_8 \) mixing angle 
(\(\theta_{\eta_1-\eta_8}\)) is \(-19.5\) degrees, and 
\(\omega_1\)-\(\omega_8 \) mixing angle is the ideal mixing 
(35 degrees). 
If we take \(a=2\) in (\ref{HLS}), we have the celebrated 
KSRF relation \(M_{\rho}^2=2f_{\pi}^2g^2\), universality of 
the \(\rho\)-meson coupling and the vector meson dominance for 
the electromagnetic form factor[4][5]. 

For obtaining the pseudoscalar meson mass terms, we introduce 
the quark mass matrix(\({\cal M}\)) as, 
\begin{equation}
{\cal L}_M = \frac{f_{\pi}^2 \mu}{2}{\rm tr\/}(\xi_R{\cal M}
\xi_L^{\dagger}+\xi_L{\cal M}\xi_R^{\dagger}) + m_{\eta_1}^2 , 
\end{equation} 
where \(\mu {\cal M}\) is relate to mass of \(\pi\), etc. and 
\(m_{\eta_1}\) is the mass term of \(\eta'\) due to \(U(1)_A\) 
breaking by gluon anomaly. 
Analogously, we may add appropriate isospin/\(SU(3)\) breaking 
terms to (\ref{HLS}) [5], 
\begin{eqnarray}
\Delta{\cal L}_{A,(V)} &=& -\frac{f_{\pi}^2}{8}{\rm tr\/} 
(D_{\mu}\xi_L \cdot \epsilon_{A,(V)}\xi_R^{\dagger} 
\pm D_{\mu}\xi_R \cdot \epsilon_{A,(V)}\xi_L^{\dagger})^2 , \\
\epsilon_{A,(V)} &=& {\rm diag\/}(0,0,\epsilon_{A,(V)}).
\end{eqnarray}

Further improvements for (\(\ref{HLS}\)) have been elaborated [10].
Here we will not discuss the non-anomalous sector (\(\ref{HLS}\)) 
any furthermore, 
because we are only interested in the anomalous sector. 
We simply assume that the parameters of the non-anomalous 
Lagrangian have been arranged so as to reproduce the relevant 
experimental data. Thus we use the experimental values as inputs 
from the non-anomalous part. 
\\
\\

In addition to (\(\ref{HLS}\)) there exists an anomalous part of 
the HLS Lagrangian. Fujiwara et al.[6] proposed how to incorporate 
vector mesons in this part of the Lagrangian without changing 
the anomaly determined by WZW term[6]. 
They have given the anomalous action as follows;
\begin{equation}
\Gamma = \Gamma_{WZW} + \sum_{i=1}^4 \int_{M^4}c_i{\cal L}_i , 
\label{WZ} 
\end{equation}
where 
\begin{eqnarray}
\Gamma_{WZW} &=& -\frac{iN_c}{240\pi^2} \int_{M^5}{\rm tr\/}
[(dU)\cdot U^{\dagger}]^5_{covariantization} \; ,  \\
{\cal L}_1 &=& {\rm tr\/} (\hat{\alpha}_L^3\hat{\alpha}_R - 
\hat{\alpha}_R^3\hat{\alpha}_L) ,  \\
{\cal L}_2 &=&{\rm tr\/} (\hat{\alpha}_L\hat{\alpha}_R\hat{\alpha}_L 
\hat{\alpha}_R) ,  \\
{\cal L}_3 &=& i {\rm tr\/} F_V(\hat{\alpha}_L \hat{\alpha}_R 
- \hat{\alpha}_R\hat{\alpha}_L) ,  \\
{\cal L}_4 &=& \frac{i}{2} {\rm tr\/} (\hat{F}_L+\hat{F}_R)\cdot
(\hat{\alpha}_L\hat{\alpha}_R-\hat{\alpha}_R\hat{\alpha}_L) , \\
\hat{\alpha}_{L,R} &=& D\xi_{L,R}\cdot\xi_{L,R}^{\dagger}=
d\xi_{L,R}\cdot \xi_{L,R}^{\dagger}-igV+ie\xi_{L,R}A
\xi_{L,R}^{\dagger} \; ,  \\
U &=& \xi_L^{\dagger}\xi_R \; ,\;  F_V = dV-igV^2,  \\
\hat{F}_{L,R} &=& \xi_{L,R}(dA - ieA^2)\xi_{L,R}^{\dagger} \: .  
\end{eqnarray}
Notice that \({\cal L}_1 \sim {\cal L}_4 \) have no contribution to 
anomalous processes such as \(\pi^0\to 2\gamma\) and 
\(\gamma\to 3\pi\) at soft momentum limit, 
because these Lagrangian are constructed with hidden-gauge covariant 
blocks such as \(\hat{\alpha}_{L,R},F_V,\hat{F}_{L,R} \)[6]. 

We take \(c_3=c_4=-15C , c_1-c_2=15C \) in (\ref{WZ}) 
for phenomenological reason[6]. 
Then we obtained the Lagrangian of anomalous sector as follows:
\begin{eqnarray}
{\cal L}_{{\rm FKTUY\/}} & = & 5C[3(VVP)-2(\gamma P^3)]+
\cdot \cdot \cdot , \label{VVP} \\
(VVP) &=& -\frac{2ig^2}{f_{\pi}} {\rm tr\/} (VdVdP+dVVdP) , 
\nonumber \\
(\gamma P^3) &=& \frac{4e}{f_{\pi}^3}{\rm tr\/}A(dP)^3 , 
\nonumber \\
C &=& -\frac{iN_c}{240\pi^2} , \nonumber 
\end{eqnarray}

Here, it is important that the amplitude such as 
\(\pi^0 \to 2\gamma, \gamma \to 3\pi\) at low energy limit are 
determined essentially only by the non-Abelian anomaly of 
the chiral \(U(3)_L  \times U(3)_R \) symmetry. 
Eq. (\(\ref{VVP}\)) is, 
of course, consistent with the low energy theorem related to 
the anomaly. 

\section{Isospin/\(SU(3)\)-breaking Terms in the Anomalous Sector}
\my
We now consider how to modify \({\cal L}_1 \sim {\cal L}_4\) 
by introducing isospin/\(SU(3)\)-breaking parameters, 
\(\epsilon\)'s, treated as ``spurions''[11]. 
The spurion \(\epsilon\) transforms as 
\(\epsilon \to g_L(x) \, \epsilon \, g_R^{\dagger}(x)\). 
Then we define the hidden-gauge covariant block 
\(\hat{\epsilon} \equiv \frac{1}{2}(\xi_L \epsilon \xi_R^{\dagger} + 
\xi_R \epsilon^{\dagger} \xi_L^{\dagger})\). 
We construct Lagrangians out of the hidden-gauge covariant blocks 
such as \(\hat{\alpha}_{L,R}\), \(F_V\), \(\hat{F}_{L,R}\) 
and \(\hat{\epsilon}\) so as to make them `` invariant " under 
\([U(3)_L  \times U(3)_R]_{{\rm global\/}} \times 
[U(3)_V]_{{\rm local\/}} \) 
as well as parity-, charge conjugation-, and \(CP\)-transformations. 
After hidden-gauge fixing, they become explicit breaking terms of 
the \(SU(3)\) symmetry. 
Then, in general, we obtain isospin/\(SU(3)\)-broken anomalous 
Lagrangians.
\begin{eqnarray}
\Delta {\cal L}_1 &=& {\rm tr\/} [\hat{\alpha}_L^3( \hat{\alpha}_R 
\cdot \hat{\epsilon}^{(1)} + \hat{\epsilon}^{(1)} \cdot 
\hat{\alpha}_R) - \hat{\alpha}_R^3(\hat{\alpha}_L \cdot 
\hat{\epsilon}^{(1)} + \hat{\epsilon}^{(1)} \cdot \hat{\alpha}_L)]  
, \\
\Delta {\cal L}'_1 &=& {\rm tr\/} (\hat{\alpha}_L 
\hat{\epsilon}^{(1')}\hat{\alpha}_L^2 \hat{\alpha}_R - 
\hat{\alpha}_R \hat{\epsilon}^{(1')}\hat{\alpha}_R^2 \hat{\alpha}_L 
+ \hat{\alpha}_L^2 \hat{\epsilon}^{(1')} \hat{\alpha}_L 
\hat{\alpha}_R   -\hat{\alpha}_R^2 \hat{\epsilon}^{(1')} 
\hat{\alpha}_R \hat{\alpha}_L),  \\
\Delta {\cal L}_2 &=&{\rm tr\/} (\hat{\epsilon}^{(2)} \cdot 
\hat{\alpha}_L + \hat{\alpha}_L \cdot \hat{\epsilon}^{(2)}) 
\hat{\alpha}_R\hat{\alpha}_L\hat{\alpha}_R , \\
\Delta {\cal L}_3 &=& i {\rm tr\/} (F_V \cdot \hat{\epsilon}^{(3)} + 
\hat{\epsilon}^{(3)} \cdot F_V) \cdot (\hat{\alpha}_L \hat{\alpha}_R 
- \hat{\alpha}_R\hat{\alpha}_L)  , \\
\Delta {\cal L}'_3 &=& i{\rm tr\/} F_V (\hat{\alpha}_L 
\hat{\epsilon}^{(3')} \hat{\alpha}_R - \hat{\alpha}_R 
\hat{\epsilon}^{(3')} \hat{\alpha}_L),  \\
\Delta {\cal L}_4 &=& i {\rm tr\/} [(\hat{F}_L \cdot 
\hat{\epsilon}^{(4)} + \hat{\epsilon}^{(4)} \cdot \hat{F}_L)
\cdot (\hat{\alpha}_L\hat{\alpha}_R -\hat{\alpha}_R \hat{\alpha}_L) 
\nonumber \\
& & + (\hat{F}_R \cdot \hat{\epsilon}^{(4)} + \hat{\epsilon}^{(4)} 
\cdot \hat{F}_R)\cdot (\hat{\alpha}_L \hat{\alpha}_R - 
\hat{\alpha}_R\hat{\alpha}_L)] , \\
\Delta {\cal L}'_4 &=& i{\rm tr\/} (\hat{F}_L+\hat{F}_R)\cdot 
(\hat{\alpha}_L \hat{\epsilon}^{(4')}
\hat{\alpha}_R - \hat{\alpha}_R \hat{\epsilon}^{(4')}\hat{\alpha}_L) 
, \\
\Delta{\cal L}_5 &=& {\rm tr\/} (\hat{\alpha}_L^2 
\hat{\epsilon}^{(5)} \hat{\alpha}_R^2 - 
\hat{\alpha}_R^2 \hat{\epsilon}^{(5)} \hat{\alpha}_L^2) ,  \\
\Delta{\cal L}_6 &=& i {\rm tr\/} (\hat{\epsilon}^{(6)}F_V - F_V 
\hat{\epsilon}^{(6)})\cdot (\hat{\alpha}_L^2- \hat{\alpha}_R^2) , \\
\Delta{\cal L}_7 &=& i {\rm tr\/} [(\hat{\epsilon}^{(7)}\hat{F}_L 
- \hat{F}_L \hat{\epsilon}^{(7)})
\hat{\alpha}_R^2 -(\hat{\epsilon}^{(7)}\hat{F}_R - \hat{F}_R 
\hat{\epsilon}^{(7)})\hat{\alpha}_L^2] , \\
\Delta{\cal L}_8 &=& i {\rm tr\/}[(\hat{\epsilon}^{(8)}\hat{F}_L 
- \hat{F}_L \hat{\epsilon}^{(8)})
 \hat{\alpha}_L^2-(\hat{\epsilon}^{(8)}\hat{F}_R - \hat{F}_R 
 \hat{\epsilon}^{(8)})\hat{\alpha}_R^2] . 
\end{eqnarray}
Here \(\hat{\alpha}_{L,R}\), \(F_V\), \(\hat{F}_{L,R}\) transform 
under \(P\) and \(C\) transformations as 
\begin{eqnarray}
P: \; \; & & \hat{\alpha}_{L,R\mu}\longrightarrow 
\hat{\alpha}_{R,L}^{\mu} \; ,\\
& & F_{V\mu\nu}\longrightarrow F_V^{\mu\nu} \: , \: 
\hat{F}_{L,R\mu\nu}\longrightarrow \hat{F}_{R,L}^{\mu\nu} \; , \\
C: \; \; & & \hat{\alpha}_{L,R}\longrightarrow -\hat{\alpha}_{R,L}^T 
\; , \\
& & F_V \longrightarrow -F_V^T \: , \: 
\hat{F}_{L,R} \longrightarrow -\hat{F}_{R,L}^T \: .
\end{eqnarray}
We could introduce \(P\)-odd ``spurion'' \(\hat{\epsilon}_- = 
\frac{1}{2}(\xi_L \epsilon \xi_R^{\dagger} - 
\xi_R \epsilon^{\dagger} \xi_L^{\dagger})\), which, however, 
is not relevant to the following analysis. 

Among the above additional terms, only 
\(\Delta{\cal L}_{3,4,6,7,8} \) terms contribute to 
the radiative decays of vector mesons. 
There still exists too many parameters. However, 
we may select the combination of \(\Delta{\cal L}_{1 \sim 8} \) 
so as to eliminate 
the direct \(V\gamma P\)-, \(VP^3\)-coupling terms, 
which do not exist in the original 
Lagrangian \({\cal L}_{{\rm FKTUY\/}}\) (\ref{VVP}). 
Then the isospin/\(SU(3)\)-broken anomalous Lagrangians consist of 
only the following two terms: 
\begin{eqnarray}
-\Delta {\cal L}_{VVP}^a &=& \frac{3g^2}{4\pi^2 f_P}
{\rm tr\/}\epsilon'(dVdVP+PdVdV) - \frac{3e^2}{4\pi^2 f_P}{\rm tr\/}
\epsilon'(dAdAP+PdAdA) \nonumber \\   
 & & + i \frac{3e}{4\pi^2 f_P^3}{\rm tr\/}\epsilon'
 (dP^3A-AdP^3+dPAdP^2-dP^2AdP), \\
-\Delta {\cal L}_{VVP}^b &=& \frac{3g^2}{2\pi^2 f_P}{\rm tr\/}
\epsilon(dVPdV) 
- \frac{3e^2}{2\pi^2 f_P}{\rm tr\/}\epsilon(dAPdA) \nonumber \\
 & & + i \frac{3e}{2\pi^2 f_P^3} {\rm tr\/}\epsilon(dP^3A-AdP^3) .  
\end{eqnarray}

Our \(\Delta {\cal L}_{VVP}^b \) resembles the \(SU(3)\)-broken 
anomalous Lagrangian introduced by Bramon et al.[8], 
but, is conceptually quite different from the latter. In fact 
the prediction on \(\eta (\eta') \to 2\gamma \) decay width 
in the latter is different from the low energy theorem's prediction. 
On the other hand, our \(\Delta{\cal L}_{VVP}^{a,b}\) do not change 
the low energy theorem by construction obviously. 

\section{Phenomenological Analysis for Radiative Decays}
\my
We now discuss the phenomenological consequences of our Lagrangian 
\({\cal L}_{{\rm anomalous\/}}={\cal L}_{{\rm FKTUY\/}} + 
\Delta {\cal L}_{VVP}^a + \Delta {\cal L}_{VVP}^b \). 
For convenience, we define relevant coupling  constant as
\begin{equation}
g_{VP\gamma}=\sum_{V'}\frac{g_{VV'P}g_{V'\gamma}}{M_{V'}^2},
\end{equation}
considering that these decays proceed via intermediate vector mesons 
\(V'\). Then we obtain each radiative decay width
\begin{eqnarray}
\Gamma(V\longrightarrow P\gamma) &=& \frac{1}{3}\alpha \cdot 
g_{VP\gamma}^2 \left( \frac{M_V^2-M_P^2}{2M_V} \right) ^3 , \\
\Gamma(\eta'\longrightarrow V\gamma) &=& \alpha \cdot 
g_{\eta' V\gamma}^2 
\left(\frac{M_{\eta'}^2 - M_V^2}{2M_{\eta'}} \right)^3 , 
\end{eqnarray} 
where \(g_{VV'P}\), \(g_{V'\gamma}\), and \(M_{V'}\) are anomalous 
\(VV'P\) coupling constant, \(V'\)-\(\gamma\) mixing, 
and mass of vector meson, respectively. 
In \(\Delta{\cal L}_{VVP}^{a,b}\) we take a parametrization 
for convenience:
\begin{equation}
\epsilon' = \left( 
\begin{array}{ccc}
-\epsilon'_1 &  &  \\
 & -\epsilon'_2 &  \\
 &  & -\epsilon'_3
\end{array}  
\right) , \;\;
\epsilon = \left( 
\begin{array}{ccc}
\epsilon_1 + \epsilon'_1 &  &  \\
 & \epsilon_2+\epsilon'_2 &  \\
 &  & \epsilon_3+\epsilon'_3
\end{array}  
\right). 
\end{equation}
Thus each \(g_{VP\gamma}\) is given in terms of the parameters in 
\(\Delta{\cal L}_{VVP}^{a,b}\) 
\begin{equation}
\left \{
\begin{array}{rcl}
g_{\rho^0\pi^0\gamma} &=& G(1+4\epsilon_1-2\epsilon_2+3\delta) ,\\
g_{\rho^{\pm}\pi^{\pm}\gamma} &=& G(1+3\epsilon'_1-3\epsilon'_2 
+4\epsilon_1-2\epsilon_2) , \\
g_{\omega\pi^0\gamma} &=& 3G(1+\frac{4}{3}\epsilon_1+
\frac{2}{3}\epsilon_2-\frac{\delta}{3}) ,\\
g_{\omega\eta\gamma}&=& \frac{f_{\pi}}{f_{\eta}}\sqrt{\frac{2}{3}}G
(1+4\epsilon_1-2\epsilon_2-\sqrt{2}\theta_V-3\delta-
\frac{\theta_P}{\sqrt{2}}) , \\
g_{\rho^0\eta\gamma} &=& \frac{f_{\pi}}{f_{\eta}}\sqrt{6}G
(1+\frac{4}{3}\epsilon_1+\frac{2}{3}\epsilon_2+\frac{\delta}{3}-
\frac{\theta_P}{\sqrt{2}}), \\
g_{\phi\eta\gamma}&=& \frac{f_{\pi}}{f_{\eta}}\frac{2}{\sqrt{3}}G
(1+2\epsilon_3+\frac{\theta_V}{\sqrt{2}}+\sqrt{2}\theta_P) , \\
g_{K^{*\pm}K^{\pm}\gamma} &=& \frac{f_{\pi}}{f_K}G(1+3\epsilon'_1
-3\epsilon'_3
+4\epsilon_1-2\epsilon_3) ,\\ 
g_{\bar{K^{*0}}\bar{K^0}\gamma} &=& -\frac{f_{\pi}}{f_K}2G
(1+\epsilon_2+\epsilon_3) , \\
g_{\phi\pi^0\gamma} &=& g_{\omega\pi^0\gamma} \cdot\theta_V , \\
g_{\eta'\rho^0\gamma} &=& \frac{f_{\pi}}{f_{\eta'}}\sqrt{3}G
(1+\frac{4}{3}\epsilon_1+\frac{2}{3}\epsilon_2+
\frac{\delta}{3}+\sqrt{2}\theta_P) , \\
g_{\eta'\omega  \gamma} &=& \frac{f_{\pi}}{f_{\eta'}}
\frac{1}{\sqrt{3}}G
(1+4\epsilon_1-2\epsilon_2+2\sqrt{2}\theta_V-
3\delta+\sqrt{2}\theta_P) , \\
g_{\phi\eta'\gamma} &=& -\frac{f_{\pi}}{f_{\eta'}}
\frac{2\sqrt{2}}{\sqrt{3}}G(1+2\epsilon_3-\frac{\theta_V}{2\sqrt{2}}
-\frac{\theta_P}{\sqrt{2}}) , \label{4-5}
\end{array}
\right.
\end{equation}
where \(G = \frac{g}{4\pi^2f_{\pi}}\). 

The parameters \(\theta_V,\theta_P\) 
appearing in the expression of \(g_{VP\gamma}\) stand for 
the difference of \(\phi\)-\(\omega\), \(\eta\)-\(\eta'\) 
mixing angles from the already determined mixing angles 
by the non-anomalous sector, i.e. , 
ideal mixing and \(\eta_1\)-\(\eta_8\) mixing, respectively. 
The parameter \(\delta\) comes from 
the \(\rho\)-\(\omega\) interference effect arising from 
the small mass difference of \(\rho\) and \(\omega\). 

2For reproducing the experimental value of 
\(\Gamma (\phi\to\rho\pi\to\pi\pi\pi)\), 
we took \(\theta_V=0.0600\pm 0.0017 \). The sign comes from 
the observed \(\phi\)-\(\omega\) interference effects in 
\(e^+e^- \to \pi^+ \pi^- \pi^0 \)[2]. 

Similarly, we consider the decay of 
\(\omega\to\pi\pi\), which is \(G\)-parity violating process. 
If the isospin were not broken, such process would not exist. 
The experimental value 
of \(\Gamma(\omega \to \pi \pi) \) is reproduced for 
\(\delta=0.0348\pm 0.0024\). 
We calculated \(\delta\) from \(\Gamma(\omega\to\pi\pi)/ 
\Gamma(\rho\to\pi\pi)=
\delta^2 \cdot \frac{p_{\omega\to\pi\pi}^3}{M_{\omega}^2}/
\frac{p_{\rho\to\pi\pi}^3}{M_{\rho}^2}\), where \(p_{V\to\pi\pi}\) 
is the final state pion momentum. 
The ambiguity of the sign has been resolved recently 
through the decays of \(\omega\) produced in 
\(\pi^- p \to \omega n\) [9], in which the constructive interference 
has been supported. 

The mixing angle \(\theta_{\eta_1-\eta_8}( =\)arcsin(\(-\)1/3)) 
has been deduced from \(\eta\)-\(\eta'\) phenomenology[12]. 
Thus we take \(\theta_P=0\). 

There are essentially five free parameters from 
\(\Delta {\cal L}_{VVP}^{a,b}\) in (\ref{4-5}), 
because \(\epsilon'_1\) is negligible. 
We search for the parameter region consistent with the experiments. 

We take \(g=4.27\pm 0.02\) from \(\Gamma(\rho \to \pi \pi)=151.2\pm 
1.2 {\rm MeV\/}\), and \(f_{\pi}=131{\rm MeV\/},f_K=160 \pm 2
{\rm MeV\/}\)[2], and \(f_{\eta}=150\pm 6{\rm MeV\/},
f_{\eta'}=142\pm 3{\rm MeV\/}\) from 
\(\eta (\eta') \to 2\gamma\) [2]. 
Then we obtained the results listed in Table 1. \\
\\

\footnotesize
\begin{table}[t]
\begin{tabular}{lrrrr} \hline
{Decay Mode} & {(I) \({\cal L}_{{\rm FKTUY\/}}\) [6]} & 
{(II) Bramon's [8] } & {(III) Ours} \qquad & {exp.[2]} \quad 
\\ \hline \hline
\(\Gamma(\rho^0\to\pi^0\gamma) \) &  \(86.2\pm 0.8{\rm keV\/}\) &  \(86.2\pm 0.8{\rm keV\/}\) &  
\(114 \pm 7 {\rm keV\/}\) &  \(121\pm 31{\rm keV\/}\) \\
\(\Gamma(\rho^{\pm}\to\pi^{\pm}\gamma) \) & \(85.6\pm 0.8{\rm keV\/}\) & \(85.6\pm 0.8{\rm keV\/}\) & \(68.1\pm 0.6{\rm keV\/}\) & \(68\pm 7{\rm keV\/}\) \\
\(\Gamma(\omega\to\pi^0\gamma ) \) &\(815\pm 8{\rm keV\/}\)& \(815\pm 8{\rm keV\/}\) & 
 \(745\pm 31{\rm keV\/}\)  &\(717\pm 43{\rm keV\/}\) \\
\(\Gamma(\omega\to\eta\gamma) \) & \(6.68\pm 0.59{\rm keV\/}\) & \(5.6\pm 0.6{\rm keV\/}\) & 
\(4.94\pm 0.75{\rm keV\/}\) & \(7.00\pm 1.77{\rm keV\/}\) \\
\(\Gamma(\rho^0\to\eta\gamma) \) & \(52.4\pm 4.6{\rm keV\/}\) & \(52.4\pm 4.6{\rm keV\/}\) & 
\(50.4\pm 6.0{\rm keV\/}\) & \(57.5\pm 10.6 {\rm keV\/}\) \\
\(\Gamma(\phi\to\eta\gamma) \) & \(80.7 \pm 7.1{\rm keV\/}\) & \(57 \pm 9{\rm keV\/}\) & 
\(62.8\pm 8.8{\rm keV\/}\) & \(56.9\pm 2.9{\rm keV\/}\) \\
\(\Gamma(K^{*\pm}\to K^{\pm}\gamma) \) & \(32.8\pm 0.9{\rm keV\/}\) & \(47\pm 5{\rm keV\/}\) &
\(50.4 \pm 1.7{\rm keV\/}\) & \(50\pm 5{\rm keV\/}\) \\ 
\(\Gamma(\bar{K^{*0}}\to\bar{K^0}\gamma) \) &\(132\pm 4{\rm keV\/}\) & \(107\pm 15{\rm keV\/}\)& \(106.9 \pm 5.5{\rm keV\/}\) & \(117\pm 10{\rm keV\/}\) \\
\(\Gamma(\phi\to\pi^0\gamma) \) & \(----\) & \(6.76\pm 0.34{\rm keV\/}\) & 
\(6.19\pm 0.55{\rm keV\/}\) & \(5.80\pm 0.58{\rm keV\/}\) \\
\(\Gamma(\eta'\to\rho^0\gamma) \) & \(61.7\pm 2.7{\rm keV\/}\) & \(61.7\pm 2.7{\rm keV\/}\) & 
\(59.3\pm 4.5{\rm keV\/}\) & \(61\pm 5{\rm keV\/}\) \\
\(\Gamma( \eta'\to\omega \gamma) \) & \(5.74\pm 0.25{\rm keV\/}\) & \(7.86 \pm 0.34{\rm keV\/}\) & \(7.14\pm 0.71 {\rm keV\/}\) & \(6.1\pm 0.8{\rm keV\/}\) \\
\(\Gamma(\phi\to\eta'\gamma) \) & \(0.827\pm 0.036{\rm keV\/}\) & \(0.5\pm 0.1{\rm keV\/}\) & 
\(0.552\pm 0.055{\rm keV\/}\) & \(<1.84{\rm keV\/}\) \\ \hline \hline
\(\Gamma(\pi^0\to 2\gamma) \) & \(7.70{\rm eV\/}\) & \(7.70{\rm eV\/}\) & \(7.70{\rm eV\/}\) & \(7.7 \pm 0.6 {\rm eV\/}\) \\
\(\Gamma(\eta\to 2\gamma) \) & \(0.46\pm 0.04{\rm keV\/}\) & \(0.51\pm 0.04{\rm keV\/}\) 
& \(0.46\pm 0.04{\rm keV\/}\) & \(0.46\pm 0.04{\rm keV\/}\) \\
\(\Gamma(\eta'\to 2\gamma) \) & \(4.26\pm 0.19{\rm keV\/}\) & \(3.6\pm 0.2{\rm keV\/}\) 
& \(4.26\pm 0.19{\rm keV\/}\) & \(4.26\pm 0.19{\rm keV\/}\) \\ \hline \hline
\(\Gamma(\eta\to \pi^+\pi^- \gamma) \) & \(0.0660\pm 0.0053{\rm keV\/}\) 
& \(0.0660\pm 0.0053{\rm keV\/}\) 
& \(0.0648\pm 0.0059{\rm keV\/}\) & \(0.0586\pm 0.0057{\rm keV\/}\) \\
\(\Gamma(\eta'\to \pi^+ \pi^- \gamma) \) & \(53.0\pm 2.2{\rm keV\/}\) 
& \(53.0\pm 2.2{\rm keV\/}\) 
& \(50.3\pm 3.4{\rm keV\/}\) & \(56.1\pm 6.4{\rm keV\/}\) \\ \hline 
\hline
\end{tabular}
\caption[table]{Radiative Decay Width of Vector Mesons }\par
\begin{center}
\begin{minipage}{0.9\textwidth}
{\scriptsize 
(I) Values of original \({\cal L}_{{\rm FKTUY\/}}\) [6] 
(II) Values of the \(SU(3)\)-broken model by Bramon et al.[8]  \\
(\(\epsilon_3 = -0.1 \pm 0.03\)). 
The predictions by this model to \(\eta (\eta') \to 2 \gamma\) 
differ from the experimental value.  \\
(III) Values of our model. The region of parameters : 
\( 0.0279 < 4 \epsilon_1-2 \epsilon_2 < 0.0670 \), \\
\( -0.0471 < \frac{4}{3} \epsilon_1 + \frac{2}{3} \epsilon_2 < -0.0174 \),
\( -0.112 < \epsilon_2 + \epsilon_3 < -0.0902 \), 
\( -0.0925 < \epsilon_3 < -0.0702 \), \\
\( 4 \epsilon_1 - 2 \epsilon_2 - 3 \epsilon'_2 = -0.108 \),
\( 4 \epsilon_1 - 2 \epsilon_3 - 3 \epsilon'_3  = 0.235 \), 
(We took the center value to \(\epsilon'\)) .
}
\end{minipage}
\end{center}
\end{table}
\normalsize

In Table 1, (I)\(\sim\)(III) mean:

\(\hspace{-1cm} \left \{
\begin{array}{l}
\mbox{\enskip (I) Values of original } \; {\cal L}_{{\rm FKTUY\/}} 
, \\
\mbox{ (II) Values of the } SU(3)
\mbox{ -broken model by Bramon et al.[8] } 
(\epsilon_3 = -0.1 \pm 0.03) \\
\qquad \mbox{ The predictions by this model to } 
\eta \; (\eta') \to 2 \gamma 
\mbox{ differ from the experimental value, } \\
\mbox{(III) Values of our model } \\
\qquad  \mbox{ The region of parameters } 
0.0279 < 4 \epsilon_1-2 \epsilon_2 < 0.0670 \; , \\
\qquad -0.0471 < \frac{4}{3} \epsilon_1 + \frac{2}{3} 
\epsilon_2 < -0.0174 
\; ,-0.112 < \epsilon_2 + \epsilon_3 < -0.0902 \; , \\
\qquad -0.0925 < \epsilon_3 < -0.0702 \; , 
 4 \epsilon_1 - 2 \epsilon_2 - 3 \epsilon'_2 = -0.108 \; ,
4 \epsilon_1 - 2 \epsilon_3 - 3 \epsilon'_3  = 0.235 \; , \\
  \qquad (\mbox{ We took the center value to } \epsilon'). 
\end{array}
\right.
\)
\\
\\

The parameter region (III) suggests that isospin/\(SU(3)\)-breaking 
effects for the anomalous sector cannot be given by the quark mass 
matrix in a simple manner. 
\\
\\

The results for \(\Gamma(\rho^0\to\pi^0\gamma),
\Gamma(\rho^{\pm}\to\pi^{\pm}\gamma),\Gamma(\omega\to\pi^0\gamma),
\Gamma(\omega\to 3\pi)\) in Table 1 
suggest that isospin breaking terms are very important. 
Both (I) and (II) in Table 1 do not have isospin breaking terms. 
These values differ substantially from the experiments, which 
cannot be absorbed by the ambiguity of the hidden-gauge coupling 
\(g\) whose value are determined either by 
\(\Gamma(\rho\to 2\pi)\) or by \(\Gamma(\rho\to e^+ e^-)\). 
In order to avoid this ambiguity, let us take some expressions 
cancelling \(g\), i.e. , \(\Gamma(\rho\to\pi\gamma)/
\Gamma(\rho\to 2\pi),\Gamma(\omega\to \pi \gamma)/
\Gamma(\rho\to 2\pi)\). 
Then we find that predictions of the original 
\({\cal L}_{{\rm FKTUY\/}}\) and Bramon et al.[8] are 
still different from the experiments. 
These Lagrangian without isospin breaking terms yields 
\begin{eqnarray}
\frac{\Gamma(\rho^0 \to\pi\gamma)}{\Gamma(\rho\to 2\pi)}
 &=& \alpha M_{\rho}^2 P_{\rho\to\pi\gamma}^3/16\pi^3 f_{\pi}^2 
 P_{\rho\to\pi\pi}^3 \; ,  \\ 
 &=& 5.6\times 10^{-4} \; , \nonumber \\
 & & [{\rm exp.\/} \; (7.9\pm 2.0) \times 10^{-4} \; ] \; , 
\nonumber \\
& & [{\rm ours \/} \; (7.5\pm 0.5) \times 10^{-4} \; ] \; , 
\nonumber \\
\frac{\Gamma(\rho^{\pm}\to\pi\gamma)}{\Gamma(\rho\to 2\pi)}
 &=& \alpha M_{\rho}^2 P_{\rho\to\pi\gamma}^3/16\pi^3 f_{\pi}^2 
 P_{\rho\to\pi\pi}^3 \; ,  \\ 
 &=& 5.6\times 10^{-4} \; , \nonumber \\
 & & [{\rm exp.\/} \; (4.5\pm 0.5) \times 10^{-4} \; ] \; , 
\nonumber \\
& & [{\rm ours \/} \; \; 4.5 \times 10^{-4} \; ] \; , \nonumber \\
\frac{\Gamma(\omega\to \pi \gamma)}{\Gamma(\rho\to 2\pi)} &=& 
9\alpha M_{\rho}^2 P_{\rho\to\pi\gamma}^3/16\pi^3 f_{\pi}^2 
P_{\rho\to\pi\pi}^3 \; ,  \\ 
&=& 5.4\times 10^{-3} \; , \nonumber \\
 & & [{\rm exp.\/} \; (4.7\pm 0.4)\times 10^{-3}] \; , \nonumber \\
& & [{\rm ours \/} \;\; (4.9\pm 0.2)\times 10^{-3}] \; . \nonumber 
\end{eqnarray}

Finally, we pay attention to 
\(\Gamma(\eta \: (\eta') \: \to \pi^+\pi^-\gamma)\), which are 
given by 
\begin{eqnarray}
\Gamma(\eta\to \pi^+\pi^- \gamma) &=& 
\frac{3 g^2 \alpha}{16\pi^6 f_{\eta}^2 M_{\eta}}
\int dE_+ dE_- [\mbox{\boldmath $p$}_+^2 \mbox{\boldmath $p$}_-^2 - 
(\mbox{\boldmath $p$}_+ \cdot \mbox{\boldmath $p$}_-)^2] \times 
\nonumber \\
 & & \left ( \frac{1+4/3\epsilon_1+2/3\epsilon_2}{(p_+ +  p_-)^2 - 
 M_{\rho}^2} 
 + \frac{1+4\epsilon_1+2\epsilon_2}{3M_{\rho}^2} \right )^2 \; , \\
\Gamma(\eta'\to \pi^+\pi^- \gamma) &=& 
\frac{3 g^2 \alpha}{32\pi^6 f_{\eta'}^2 M_{\eta'} }
\int dE_+ dE_- [\mbox{\boldmath $p$}_+^2 \mbox{\boldmath $p$}_-^2 - 
(\mbox{\boldmath $p$}_+ \cdot \mbox{\boldmath $p$}_-)^2] \times 
\nonumber \\
 & & \left ( \frac{1+4/3\epsilon_1+2/3\epsilon_2}{(p_+ +  p_-)^2 +
iM_{\rho} \Gamma_{\rho}- M_{\rho}^2} + 
\frac{1+4\epsilon_1+2\epsilon_2}{3M_{\rho}^2} \right )^2 \; , \\
\Gamma_{\rho} &=& \Gamma(\rho\to 2\pi) \cdot \left( 
\frac{(q_{\pi^+}+q_{\pi^-})^2-4M_{\pi}^2}{M_{\rho}^2-4M_{\pi}^2} 
\right) ^{3/2} 
\theta(\: (q_{\pi^+}+q_{\pi^-})^2-4M_{\pi}^2 \: ) \; , \nonumber 
\end{eqnarray}
where we expressed \(\rho\)-meson propagater in the process 
\(\eta' \to \rho^0 \gamma \to \pi^+ \pi^- \gamma \) 
by using the decay width of the \(\rho\)-meson \(\Gamma_{\rho}\). 
\\

\section{Hadronic Anomalous Decays}
\my
In this section, we consider hadronic anomalous decays such as 
\(\Gamma(\omega\to 3\pi)\). However, the experimental value is 
presently available only for \(\Gamma(\omega\to 3\pi)\). 
As the previous section, we obtained Table 2 for the models of 
(I)\(\sim\)(III).

\footnotesize
\begin{table}[t]
\begin{tabular}{lrrrr} \hline
{Decay Mode} & {(I) \({\cal L}_{{\rm FKTUY\/}}\) [6]}  & 
{(II) Bramon's [8]}\quad & 
{(III) Ours} \qquad & {exp. [2]} \quad \\ \hline \hline
\(\Gamma(\omega\to \pi^0 \pi^+ \pi^-) \) & \(8.18\pm 0.23{\rm MeV\/}\) & \(8.18\pm 0.23{\rm MeV\/}\) & \(7.62\pm 0.26{\rm MeV\/}\) & \(7.49\pm 0.12{\rm MeV\/}\) \\ \hline
\(\Gamma(\rho^0 \to \pi^0 \pi^+ \pi^-) \) & \(----\) & \(----\) & \(6.70\pm 3.25 {\rm keV\/}\) & \(<18{\rm keV\/}\) \\
\(\Gamma(\rho^{\pm} \to \pi^{\pm} \pi^0 \pi^0) \) & \(----\) & \(----\) & \(4.76\pm 1.70{\rm keV\/}\) & \(-----\) \\
\(\Gamma(\rho^{\pm} \to \pi^{\pm} \pi^+ \pi^-) \) & \(----\) & \(----\) & \(0.125 \pm 0.121{\rm keV\/}\) & \(-----\) \\
\(\Gamma(K^{*-} \to \bar{K^0} \pi^0 \pi^-) \) & \(17.9\pm 1.6{\rm keV\/}\) & \(14\pm 2{\rm keV\/}\) & \(12.3\pm 1.1{\rm keV\/}\) & \(<35{\rm keV\/}\) \\
\(\Gamma(K^{*-} \to K^- \pi^+ \pi^-) \) & \(8.65\pm 0.33{\rm keV\/}\) & \(6.6\pm 0.9{\rm keV\/}\) & \(6.26\pm 0.55{\rm keV\/}\) & \(<40{\rm keV\/}\) \\
\(\Gamma(K^{*-} \to K^- \pi^0 \pi^0) \) & \(1.11 \pm 0.04{\rm keV\/}\) & \(0.72\pm 0.06{\rm keV\/}\) & \(0.685 \pm 0.081{\rm keV\/}\) & \(-----\) \\
\(\Gamma(\bar{K^{*0}} \to K^- \pi^0 \pi^+) \) & \(23.2 \pm 2.1{\rm keV\/}\) & \(18\pm 3{\rm keV\/}\) & \(16.8\pm 1.6{\rm keV\/}\) & \(-----\) \\
\(\Gamma(\bar{K^{*0}} \to \bar{K^0} \pi^- \pi^+) \) & \(9.04\pm 0.83{\rm keV\/}\) & \(7.0\pm 0.1{\rm keV\/}\) & \(6.26\pm 0.60{\rm keV\/}\) & \(< 35{\rm keV\/}\) \\
\(\Gamma(\bar{K^{*0}} \to \bar{K^0} \pi^0 \pi^0) \) & \(1.11\pm 0.05{\rm keV\/}\) & \(0.71\pm 0.06{\rm keV\/}\) & \(0.522\pm 0.05{\rm keV\/}\) & \(-----\) \\
\hline \hline
\end{tabular}
\caption[table]{Hadronic Decay Width of Vector Mesons} 
\begin{center}
\begin{minipage}{0.8\textwidth}
{\scriptsize 
(I) Values of original \({\cal L}_{{\rm FKTUY\/}}\)  
(II) Values of the \(SU(3)\)-broken model by Bramon et al.[8] \\
(\(\epsilon_3 = -0.1 \pm 0.03\))  
(III) Values of our model. The region of parameters : \\
\( 0.0279 < 4 \epsilon_1-2 \epsilon_2 < 0.0670 \), 
\( -0.0471 < \frac{4}{3} \epsilon_1 + \frac{2}{3} \epsilon_2 < -0.0174 \),
\( -0.112 < \epsilon_2 + \epsilon_3 < -0.0902 \), 
\( -0.0925 < \epsilon_3 < -0.0702 \),
\( 4 \epsilon_1 - 2 \epsilon_2 - 3 \epsilon'_2 = -0.108 \),
\( 4 \epsilon_1 - 2 \epsilon_3 - 3 \epsilon'_3  = 0.235 \),\\
(We took the center value to \(\epsilon'\)).
}
\end{minipage}
\end{center}
\end{table}

\normalsize

In Table 2, we took \(K^* K \pi\)-coupling 
as about \(1.05 \times g_{K^* K \pi}\), which is given by 
(\ref{HLS}), considering 
\(\Gamma(K^{*\pm} \to (K \pi)^{\pm})=49.8 \pm 0.8\) MeV, 
\(\Gamma(K^{*0} \to (K \pi)^0)=50.5 \pm 0.6\) MeV. 

As to \(\Gamma(\omega\to 3\pi)\), we would have to consider 
the effects of \((VP^3)\)-terms from \(\Delta{\cal L}_{1,2,5} \). 
But their contributions seem to be very small compared with the 
contributions from \(\Delta{\cal L}_{VVP}^{a,b}\), 
because our prediction of \(\Gamma(\omega\to 3\pi)\) is already 
consistent with the experimental value. Therefore, it is sufficient 
to introduce the isospin/\(SU(3)\)-breaking terms only for 
\(\Delta{\cal L}_{VVP}^{a,b}\) . 
\\
\\

In Table 2 it is again suggested 
that isospin breaking terms are very important. 
As in the previous section, 
let us take some expressions 
cancelling \(g\), i.e. , \(\Gamma(\omega\to 3\pi)/
\Gamma(\rho\to 2\pi)^3\). 
Then we find that predictions of the original 
\({\cal L}_{{\rm FKTUY\/}}\) 
and Bramon et al.[8] are again different 
from the experiments. 
\begin{eqnarray}
\frac{\Gamma(\omega\to 3\pi)}{\Gamma(\rho\to 2\pi)^3} &=& 
\frac{81M_{\omega} M_{\rho}^6 (1+\epsilon_1+\epsilon_2)^2}
{256\pi^4 f_{\pi}^2 P_{\rho\to\pi\pi}^9 }\int dE_+ dE_- 
[\mbox{\boldmath $p$}_+^2 \mbox{\boldmath $p$}_-^2 - 
(\mbox{\boldmath $p$}_+ \cdot \mbox{\boldmath $p$}_-)^2]\times 
\nonumber \\
 & & \hspace{-1cm} (\frac{1}{(p_0 + p_+)^2 + M_{\rho}^2}+
 \frac{1}{(p_+ +  p_-)^2+ M_{\rho}^2}+
 \frac{1}{(p_- + p_0)^2+M_{\rho}^2})^2 \\
\nonumber \\
& =& 2.38\times 10^{-6} \; {\rm MeV\/}^{-2} 
\mbox{ ( from (I) and (II) with }
\epsilon_1 = \epsilon_2 = 0 \; ) \nonumber \\
 & & [ \; (2.16\pm 0.09)\times 10^{-6} \; {\rm MeV\/}^{-2} \; 
\mbox{ from exp. } \;] \nonumber \\
& & [ \; (2.25\pm 0.06)\times 10^{-6} \; {\rm MeV\/}^{-2} \; 
\mbox{ from (III) Ours } ] \nonumber 
\end{eqnarray}

Although the only upper bound of \(\Gamma(\rho\to 3\pi)\) and 
\(\Gamma(K^*\to K\pi\pi)\) is available now, 
it will be interesting that their value will be determined 
by the experiments in future. 

\section{Summary}
By introducing isospin/\(SU(3)\)-broken 
\(\Delta{\cal L}_{VVP}^{a,b}\) with a few parameters, 
we have shown that each decay width of anomalous process can 
be consistently reproduced with all the experimental data. 

We also made predictions: 

\begin{table}[h]
\begin{center}
\begin{tabular}{rclr} \hline
{ Decay Mode } & & { Width (keV)} & { Branching Ratio } \\ \hline \hline
\(\Gamma(\rho^0\to\pi^0\gamma)\) &=& \(114 \pm 7 \) & \((7.54 \pm 0.47)\times 10^{-4}\) \\
\(\Gamma(\phi\to\eta'\gamma)\) &=& \(0.552\pm 0.055\) & 
\((1.25 \pm 0.13) \times 10^{-4}\)  \\ 
\(\Gamma(\rho^0 \to \pi^0 \pi^+ \pi^-)\) &=& \(6.70\pm 3.25\) & \((4.43\pm 2.15)\times 10^{-5}\)  \\
\(\Gamma(\rho^{\pm} \to \pi^{\pm} \pi^0 \pi^0)\) &=& \(4.76\pm 1.70\) & \(
(3.15 \pm 1.12) \times 10^{-5}\)  \\
\(\Gamma(\rho^{\pm} \to \pi^{\pm} \pi^+ \pi^-)\) &=& \(0.125 \pm 0.121\) & \(
(8.28 \pm 8.00) \times 10^{-7}\)  \\
\(\Gamma(K^{*-} \to \bar{K^0} \pi^0 \pi^-)\) &=& \(12.3\pm 1.1\) & \(
(2.47\pm 0.22) \times 10^{-4}\)  \\
\(\Gamma(K^{*-} \to K^- \pi^+ \pi^-)\) &=& \(6.26\pm 0.55\) & \(
(1.26 \pm 0.11) \times 10^{-4}\)  \\
\(\Gamma(K^{*-} \to K^- \pi^0 \pi^0)\) &=&  \(0.685 \pm 0.081\) & \(
(1.38 \pm 0.16) \times 10^{-5}\)  \\
\(\Gamma(\bar{K^{*0}} \to K^- \pi^0 \pi^+)\) &=& \(16.8\pm 1.6\) & \(
(3.33 \pm 0.32) \times 10^{-4}\)   \\
\(\Gamma(\bar{K^{*0}} \to \bar{K^0} \pi^- \pi^+)\) &=& \(6.26\pm 0.60\) & \(
(1.24 \pm 0.12) \times 10^{-4}\)   \\
\(\Gamma(\bar{K^{*0}} \to \bar{K^0} \pi^0 \pi^0)\) &=& \(0.522\pm 0.05\) & \(
(1.03 \pm 0.10) \times 10^{-5}\)  \\ \hline 
\end{tabular}
\caption[table]{The List of Our Predictions}
\end{center}
\end{table}

We expect that the decay data for pseudoscalar mesons and 
vector mesons, such as 
\(\phi\to\eta'\gamma, \rho^0\to\pi^0\gamma\) etc., 
will be obtained with good accuracy in 
the DA\(\Phi\)NE \(\phi\)-factory. 

\section*{ \hspace{4cm} Acknowledgements}

The author is very grateful to K. Yamawaki for suggesting 
this subject, helpful discussions and also for careful reading 
the manuscript. 
Thanks are also due to M. Sugiura, A. Shibata and 
others in the laboratory at Nagoya University 
for discussions and encouragement.

\section*{[Note added]}

After completion of this work, the author was informed by 
M. Harada that a similar analysis has been done by M. Harada and 
J. Schechter who, however, assumed the breaking term is 
proportional to the quark mass matrix and 
thereby arrived at inconsistency with the data.

\end{document}